\documentclass[twocolumn,preprintnumbers,superscriptaddress,nofootinbib,aps,prl,floatfix]{revtex4}

\usepackage{amsmath,amssymb}
\usepackage{dsfont}
\usepackage{graphicx} %loads the graphicx.sty package 
\usepackage{epstopdf} %loads the epstopdf.sty package 
\usepackage{slashed}
\usepackage{subfigure}
\usepackage{color}
\usepackage{multirow}
\usepackage{hyperref}

\usepackage{footmisc}

\usepackage{array}
\newcolumntype{L}[1]{>{\raggedright\let\newline\\\arraybackslash\hspace{0pt}}m{#1}}
\newcolumntype{C}[1]{>{\centering\let\newline\\\arraybackslash\hspace{0pt}}m{#1}}
\newcolumntype{R}[1]{>{\raggedleft\let\newline\\\arraybackslash\hspace{0pt}}m{#1}}

\newcommand{\be}{\begin{eqnarray*}}
\newcommand{\ee}{\end{eqnarray*}}
\newcommand{\gl}[1]{(\ref{#1})}

\newcommand{\bee}{\begin{eqnarray}}
\newcommand{\eee}{\end{eqnarray}}
\newcommand{\beeq}{\begin{equation}}
\newcommand{\eeeq}{\end{equation}}

\newcommand{\BR}{{\text{BR}}}

\begin{document}

\title{Further on up the road: $hhjj$ production at the LHC}

\begin{abstract}
  A measurement of the $hh+2j$ channel at the LHC would be
  particularly thrilling for electroweak physics. It is not only the
  leading process which is sensitive to the $W^+W^- hh$ and $ZZhh$
  interactions, but also provides a potentially clear window to study
  the electroweak symmetry-breaking sector by probing Higgs-Goldstone
  interactions through the weak boson fusion component of the
  scattering process. Until now, a phenomenologically complete
  analysis of this channel has been missing. This is mostly due to the
  high complexity of the involved one-loop gluon fusion contribution
  and the fact that a reliable estimate thereof cannot be obtained
  through simplified calculations in the $m_t\to \infty$ limit. In
  particular, the extraction of the Higgs trilinear coupling from this
  final state rests on a delicate $m_t$-dependent interference pattern
  which is not captured in an effective field theory approach. In this
  paper, we report on the LHC's potential to access di-Higgs
  production in association with two jets in a fully-showered
  hadron-level analysis. Our study includes the finite top and bottom
  mass dependencies for the gluon fusion contribution. On the basis of
  these results, we also comment on the potential sensitivity to the
  trilinear Higgs and $VV^\dagger hh$ ($V=W^\pm,Z$) couplings that can
  be expected from measurements of this final state.
\end{abstract}
\author{Matthew J. Dolan} %\email{m.j.dolan@durham.ac.uk}
\affiliation{Theory Group, SLAC National Accelerator Laboratory,\\Menlo Park, CA 94025, USA\\[0.1cm]}
\author{Christoph Englert} %\email{christoph.englert@glasgow.ac.uk}
\affiliation{SUPA, School of Physics and Astronomy, University of
  Glasgow,\\Glasgow, G12 8QQ, United Kingdom\\[0.1cm]}
\author{Nicolas Greiner} %\email{greiner@mpp.mpg.de}
\affiliation{Max-Planck-Institut f\"ur
  Physik,\\ F\"ohringer Ring 6, 80805 M\"unchen, Germany\\[0.1cm]}
\author{Michael Spannowsky} %\email{michael.spannowsky@durham.ac.uk}
\affiliation{Institute for Particle Physics Phenomenology, Department
  of Physics,\\Durham University, DH1 3LE, United Kingdom\\[0.1cm]}

\preprint{IPPP/13/80, DCPT/13/160, SLAC-PUB-15752, MPP-2013-275}

\maketitle

%\section{Context and Introductory Remarks}
\noindent \underline{\emph{Context and Introductory Remarks.}}  After
the recent discovery of the Higgs boson \cite{orig} at the LHC
\cite{Chatrchyan:2012ufa} and subsequent analyses of its
interactions with known matter~\cite{Hcoup}, a coarse-grained picture
of consistency with the Standard Model (SM) expectation appears to be
emerging. Taking this agreement at the level of Higgs couplings at
face value, an immediate question that comes to mind is whether the
realisation of electroweak symmetry breaking (EWSB) is in fact SM-like
too.

Crucial to Higgs sector-induced EWSB is the presence of higher order
monomials of the Higgs field in the potential which misalign the Higgs
field from the electroweak symmetry-preserving direction, thus
realising the gauge symmetry non-linearly via the Higgs
mechanism. These terms are currently unknown and it is experimentally unclear whether
they exist at all.

In the minimal approach of the SM the potential reads
\begin{multline}
  \label{eq:higgspot}
  V(H^\dagger H) =
  \mu^2 H^\dagger H + \eta (H^\dagger H)^2 \\
  \leadsto {1\over 2} m_h^2h^2 + \sqrt{ {\eta\over 2}}m_h h^3 +
  {\eta\over 4}h^4\,
\end{multline}
(after the Higgs doublet field is expanded around its vacuum
expectation value (vev) in unitary gauge). Therefore, in the
SM context, the quartic and trilinear Higgs couplings are directly
related to the Higgs pole mass, the vev as set by, {\it e.g.}, the $W$
mass, and the electroweak coupling.

When reconstructing the symmetry-breaking Higgs potential in a fully
model-independent way by measuring the coefficients of the potential's Taylor
expansion about the symmetry breaking minimum $\left\langle H\right
\rangle$, the impact of having finally observed the $125$~GeV boson is
rather limited. Discovering the Higgs with SM-compatible $W$ and $Z$
couplings does not provide any additional information other than the
mere existence of a symmetry breaking vacuum (basically already known
from observing massive gauge bosons) and the size of the curvature of
the potential around the local minimum.\footnote{Interpreting the 125
  GeV boson as a pseudo-dilation is formally the only exception to
  this argument with, however, typically little theoretical appeal and
  predictivity.} These are rather generic symmetry-breaking
properties. Indeed, models which interpret the Higgs field as an
iso-doublet of \hbox{(pseudo-)}Nambu-Goldstone fields~\cite{silh}
generally involve Higgs self interactions significantly different from
the SM. This typically has far-reaching phenomenological consequences
for multi-Higgs production~depending on the involved couplings and the
presence of additional (often fermionic)
matter~\cite{thaler,contino,furlan}. While details inevitably depend
on the particular model, a measurement of the Higgs potential
coefficient $\propto h^3$ undoubtedly sheds light on the source of
EWSB by measuring the first quantity that reflects the dynamics
of~$V$. As a consequence, any knowledge about the trilinear Higgs
coupling can be used to discriminate between various realisations of
EWSB.

This has been the main motivation to study the LHC's potential to
reconstruct the Higgs trilinear coupling
$\lambda_{\text{SM}}=\sqrt{\eta/2}\,m_h$ through measuring di-Higgs
production cross sections~\cite{nigel,uli}.\footnote{A measurement of
  the quartic Higgs interactions from triple Higgs final states
  appears impossible due to the tiny signal cross
  section~\cite{rauch}.}  Early analyses have revealed sensitivity to
di-Higgs production in rare decays $hh\to b\bar b \gamma
\gamma$~\cite{tilman}, which has been reviewed by ATLAS in
Ref.~\cite{atlashh} only recently. The signal yield, however, is too
small to tightly constrain the Higgs trilinear coupling in this
channel.  Hence, given the small production cross section of inclusive
di-Higgs production at the LHC, it is imperative to apply and push
state-of-the-art reconstruction and background rejection techniques
for di-Higgs final states. For instance, the application of boosted
$h\to b\bar b$ reconstruction techniques as discussed in
Ref.~\cite{htagging} and used in $pp\to hZ+X$ analyses \cite{boostexi}
has also revealed a potentially large sensitivity to the trilinear
coupling in the $b\bar b \tau^+\tau^-$ final
states~\cite{us,us2}.\footnote{Since then, analyses of (boosted)
  di-Higgs final states have gained at a lot of interested in the
  context of the SM and
  beyond~\hbox{\cite{Papaefstathiou:2012qe,bsm,qcd}}. The Higgs boost
  is crucial to a successful analysis in the $b\bar b\tau^+\tau^-$
  channel \cite{us2}, inclusive analysis are dominated by the $t\bar
  t$ background~\cite{Baur:2003gpa}.} In addition to new analysis
strategies focussing on diverse phase space regions, it is also
mandatory to extend the list of available hadron collider processes
which can be included into a combined limit across various channels.

A process along this line which is also of outstanding theoretical
relevance is the production of a Higgs pair in association with two
jets via weak boson fusion (WBF). This contribution to $pp\to hhjj+X$
production at the LHC is particularly interesting because the WBF
component involves the quartic $VV^\dagger hh$ ($V=W^\pm ,Z$) vertices
with couplings $g_{WWhh}=e^2/(2s_w^2), g_{ZZhh}=e^2/(2s_w^2c_w^2)$.
The process obviously shares the QCD properties of single Higgs
production via WBF, making (higher-order QCD) calculations
straightforwardly adaptable from the latter
process~\cite{hhjj1,hhjj2}.  However, a comprehensive signal
vs. background investigation of the $hhjj$ final state and an analysis
of the expected sensitivity to the $VV^\dagger hh$ and trilinear Higgs
couplings have not been performed so far. The purpose of the present
work is to provide a first step in this direction.

One of the reasons for the lack of phenomenologically complete studies
of this particular final state (apart from experimental issues that we
are not qualified to comment on) is the highly involved modelling and
up to now unknown size of the gluon fusion (GF) contribution to $pp\to
hhjj+X$ at ${\cal{O}}(\alpha_s^4\alpha^2)$.\footnote{The similarity of
  the WBF component with single Higgs production via WBF allows us to
  neglect interference effects of the two signal
  contributions~\cite{Bredenstein:2008tm}.} On the one hand, the
straightforward application of low-energy effective theorems to
gluon-Higgs interactions~\cite{kniehl}
\begin{equation}
  \label{eq:heff}
  {\cal{L}_{\mathrm{eff}}} = {1 \over 4}{\alpha_s \over
    3\pi} G^a_{\mu\nu} G^{a\, \mu\nu} \log(1+ h/v)\, ,
\end{equation}
is fairly simple. On the other hand, however, since momentum transfers
$p_{T,h}~\sim m_t$ probe the kinematic region where interference with
the Higgs trilinear diagrams becomes relevant for the integrated cross
section~\cite{uli2,us}, integrating out the top quark cannot be
justified in phenomenological investigations. Destructive interference
of the effective $gghh$ and $ggh$ vertices is encoded in the log's
expansion of Eq.~\gl{eq:heff}, yet all kinematic information is lost
when employing the limit $m_t\to \infty$. Therefore, a targeted and
reliable phenomenological analysis of the di-Higgs final state must
not be based on Eq.~\gl{eq:heff}.

Keeping the full quark mass dependencies in the gluon fusion component
of $pp\to hh jj$ is a computationally intense task at the frontier of
one-loop multi-leg calculations. Given the high complexity of this
process we obtain a calculation time of up to $\sim 1$ minute per
phase space point and per massive fermion in the loop for the pure
gluon case $gg\to hhgg$, which exhibits the largest complexity with
around one thousand diagrams (for details see below). Clearly,
traditional Monte-Carlo event generation approaches do not promise a
successful outcome unless the calculation time is significantly
improved. In the following we choose to perform a phase space
point-dependent re-weighting of the effective theory to overcome this
predicament. This allows us to provide a first analysis of the $hhjj$
final state at the LHC. We also comment on the influence of
modifications of the Higgs trilinear and $VV^\dagger hh$ couplings on
the resulting $pp\to hhjj+X$ phenomenology.  
\vskip1\baselineskip

%\section{Elements of the analysis}
\noindent\underline{\emph{Elements of the Analysis.}}
An apparent difference compared to single Higgs production studies in
the two-jet category is the small cross section that is expected for
$pp\to hhjj+X$ of inclusive ${\cal{O}}(10~{\text{fb}})$. Typical GF
and QCD background suppression tools for a $125$ GeV Higgs boson such
as {\it e.g.} a central jet veto (CJV) are not applicable because in
order to observe a signal in the first place we have to rely on large
Higgs branching ratios to bottom quarks ($\BR(h\to b\bar b)\simeq
58\%$), hadronically decaying $W$s ($\BR(h\to W^+W^-)\simeq 22\%$),
and tau leptons ($\BR(h\to \tau^+\tau^-) \simeq 6\%$). All these decay
modes give rise to hadronic activity in the central detector
region. (Semi-)leptonic $Z$ boson decays are too limited by small
branching ratios ($\BR(h\to ZZ)\simeq 2.1\%$, $\BR(Z\to
e^+e^-\oplus\,\mu^+\mu^-)\simeq 7\%$ and $\BR(Z\to
\text{hadrons})\simeq 70\%$) to be of any phenomenological relevance
in this case; e.g. the fully leptonic Higgs decay is clean but
extremely rare $\BR(h\to 4\ell)\simeq 7\times
10^{-5}$. Hadronically-decaying $W$s from a Higgs decay as considered
in Ref.~\cite{Papaefstathiou:2012qe} rely on extreme boosts of the
recoiling Higgs decaying to bottoms and a resolution of the latter
decay at a level comparable to the detector granularity.  Such an
approach does not seem promising in the present case due to the small
cross section of $hhjj$ production and the larger systematic
uncertainties of the multi-jet final state. Relaxing the CJV criterion
in favour of a large invariant mass cut on the tagging jets
\cite{DelDuca:2006hk} is insufficient to tame the background
contributions and is troubled by large combinatorial uncertainties and
small statistics (see below).\footnote{However, it might be able to
  compensate this by folding in matrix elements to the analysis,
  generalizing the approach of Ref.~\cite{Andersen:2012kn}.} The most
promising avenue is therefore a generalisation of the boosted final
state analysis of~Ref.~\cite{us} to a lower $p_T$ two-jet category: On
the one hand, the signal cross section remains large by focussing on
the $hh\to b\bar b \tau^+\tau^-$ final state and combinatorial issues
can be avoided (i.e. through boosted kinematics and substructure
techniques).

We generate signal events with {\sc{MadEvent}} v4~\cite{Alwall:2007st}
and v5~\cite{Alwall:2011uj} for the WBF and GF contributions,
respectively. The former event generation includes a straightforward
add-on that allows to include the effect of modified Higgs trilinear
coupling. The GF event generation employs the
{\sc{FeynRules}}/{\sc{Ufo}}~\cite{Degrande:2011ua} tool chain to
implement the higher dimensional operators relevant for GF-induced
$hhjj$ production in the $m_t\to \infty$ limit. We pass the events to
{\sc{Herwig++}}~\cite{Bahr:2008pv} for showering and
hadronisation. For background samples we use
{\sc{Sherpa}}~\cite{Gleisberg:2008ta} and {\sc{MadEvent}}~v5,
considering $tth$, $t\bar{t}jj$, $ZWWjj$, $ZHjj$ and $ZZjj$. As in the
$hh$ and $hhj$ cases the dominant background is due to $t\bar{t}$.  We
normalise the background samples using NLO $K$-factors, namely
$0.611~\mathrm{pb}$ for $tth$~\cite{tthnlo}, $300.5~\mathrm{pb}$ for
$t\bar{t}jj$~\cite{ttjjnlo}. We adopt the a total flat $K$ factor of
$1.2$ for $Zh+2j$ motivated from Ref.~\cite{ZHnlo}. We have checked
that all other backgrounds are completely negligible. The QCD
corrections for the signal are known to be small for the WBF
contribution \cite{hhjj1,hhjj2}. It is reasonable to expect that the
corrections for the GF contributions will be similar to the $pp\to
hjj$ following the arguments of Ref.~\cite{Campbell:2006xx}, however,
we choose to remain conservative and do not include a NLO $K$ factor
guess for the GF contribution.

We correct the deficiencies of the GF event generation in the $m_t\to
\infty $ limit via an in-house re-weighting library which is called at
runtime of the analysis. We include the effects of finite top and
bottom quark masses, which are treated as complex parameters. The
value of the Higgs trilinear coupling can be steered externally. For
the generation of the matrix elements we used {\sc{GoSam}}
\cite{gosam}, a publicly available package for the automated
generation of one-loop amplitudes. It is based on a Feynman
diagrammatic approach using {\sc{QGRAF}}~\cite{qgraf} and
{\sc{FORM}}~\cite{form} for the diagram generation, and
{\sc{Spinney}}~\cite{spinney}, {\sc{Haggies}}~\cite{haggies} and
{\sc{FORM}} to write an optimised fortran output. The reduction of the
one-loop amplitudes was done using {\sc{Samurai}}~\cite{samurai},
which uses a $d$-dimensional integrand level decomposition based on
unitarity methods~\cite{unitarity}. The remaining scalar integrals
have been evaluated using {\sc{OneLoop}}~\cite{oneloop}.
Alternatively, {\sc {GoSam}} offers a reduction based on tensorial
decomposition as contained in the {\sc{Golem95}}
library~\cite{golem95}.The {\sc GoSam} framework has been used recently
for the calculation of signal and background processes important 
for Higgs searches at the LHC ~\cite{gosamref}.

%%%%%%%%%%%%%%%%%%%%%%%%%%%%%%%%%%%%%%%%%%%%%%%%%%%%%%%%%%
\begin{figure}[!b]
  \includegraphics[width=0.43\textwidth]{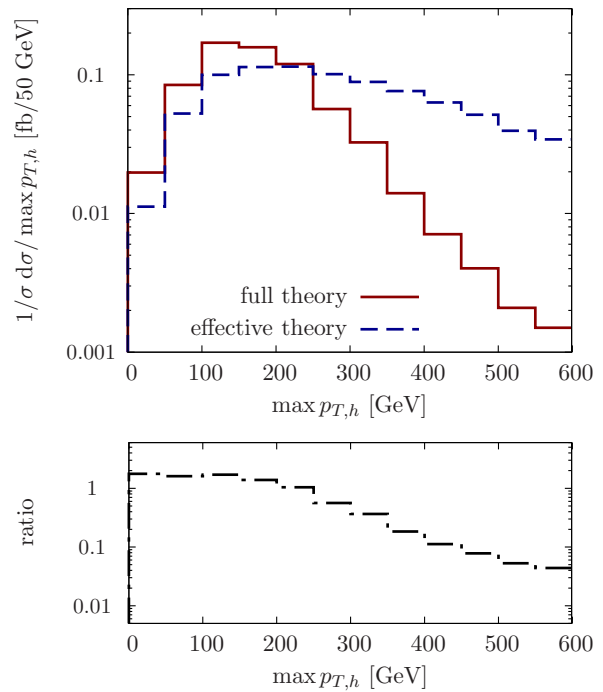}
  \caption{\label{fig:re-weight} $\max p_{T,h} $ distribution and
    effective theory vs. full theory comparison as a function of the
    maximum Higgs transverse momentum of the fully showered and
    hadronised gluon fusion sample (satisfying the parton-level
    generator cuts $p_{T,j}\geq 20$~GeV and $|\eta_j|<4.5$).}
\end{figure}
%%%%%%%%%%%%%%%%%%%%%%%%%%%%%%%%%%%%%%%%%%%%%%%%%%%%%%%%%%

%%%%%%%%%%%%%%%%%%%%%%%%%%%%%%%%%%%%%%%%%%%%%%%%%%%%%%%%%%
\begin{table*}[!t]
\renewcommand\arraystretch{1.3}
\begin{tabular}{|| l | C{2cm} |  C{2cm} | C{2cm} | C{2cm} | C{2cm} |  C{2cm} ||}
\hline
& \multicolumn{3}{ c |}{Signal with $\xi\times \lambda$} &
\multicolumn{2}{ c |}{Background} & $S/B$ \\
 & $\xi=0$ & $ \xi=1$ & $\xi=2$ & $t\bar{t}jj$ & Other BG & ratio to $\xi=1$ \\ \hline
%before cuts & &&&& & \\ 
%WBF trigger cuts & 1.436 & 0.669  & 0.693  & 3339.0 & 615.8 &
%$0.155\times 10^{-3}$ \\ 
tau selection cuts & 0.212 & 0.091 & 0.100 & 3101.0& 57.06  & $0.026\times 10^{-3}$\\ 
Higgs rec. from taus & 0.212 & 0.091  & 0.100 &  683.5 & 31.92 & $0.115\times 10^{-3}$\\ 
Higgs rec. from $b$ jets & 0.041 & 0.016 & 0.017  & 7.444   & 0.303 & $1.82\times 10^{-3}$\\ 
2 tag jets  & 0.024& 0.010  & 0.012  & 5.284&  0.236 & $1.65\times 10^{-3}$\\ 
\hline
incl. GF after cuts/re-weighting & 0.181 & 0.099 & 0.067 &  5.284&  0.236  &
$ 1 / 61.76$\\
\hline 
\end{tabular}
\vskip 0.5cm
\begin{tabular}{|| l | C{2cm} |  C{2cm} | C{2cm} | C{2cm} | C{2cm} ||}
\hline
& \multicolumn{3}{ c |}{Signal with $\zeta \times \{g_{WWhh},g_{ZZhh}\}$} &
\multicolumn{2}{ c ||}{Background}  \\
 & $\zeta=0$ & $ \zeta=1$ & $\zeta=2$ & $t\bar{t}jj$ & Other BG \\
\hline
%trigger cuts & 8.664  &  0.669    & 5.123 & 3339.0 & 615.8  \\ 
tau selection cuts & 1.353 & 0.091  & 0.841 & 3101.0& 57.06   \\ 
Higgs rec. from taus & 1.352 & 0.091  &  0.840 &  683.5 & 31.92   \\ 
Higgs rec. from $b$ jets & 0.321 & 0.016 & 0.207  & 7.444   & 0.303  \\ 
2 tag jets/re-weighting  & 0.184 & 0.010 & 0.126 & 5.284&  0.236  \\ 
\hline
incl. GF after cuts/re-weighting & 0.273 &  0.099 & 0.214 & 5.284 & 0.236 \\
\hline 
\end{tabular}
\caption{\label{tab:analysis} Cross sections in fb of the hadron-level
  analysis described in the text, including results with modified
  Higgs trilinear and $VV^\dagger hh$ couplings. Signal cross sections
  already include the branching ratios to the $h\to b\bar
  b,\tau^+\tau^-$ final states. The top four rows refer to the WBF
  sample and the last line includes the re-weighted GF
  contribution. For details see text.}
\end{table*}
%%%%%%%%%%%%%%%%%%%%%%%%%%%%%%%%%%%%%%%%%%%%%%%%%%%%%%%%%%

The maximum transverse momentum of the Higgs bosons is a good variable
to compare effective with full theory. For inclusive $hhjj$ production
we find a re-weighted distribution as depicted in
Fig.~\ref{fig:re-weight}. Qualitatively, the re-weighting pattern
follows the behaviour anticipated from $pp\to hhj$ production
\cite{us} and $pp\to hjj$~\cite{vbfnlo,DelDuca:2006hk}. As expected,
the shortcomings of the effective calculation for double Higgs
production are more pronounced than for single Higgs production:
Already for low momentum transfers the effective theory deviates from
the full theory by factors of two, making the correction relevant even
for low momenta, where one might expect the effective theory to be in
reasonably good shape. It is precisely the competing and
$m_t$-dependent contributions alluded to earlier which are not
reflected in the effective theory causing this deviation. When the
effective operators are probed at larger momentum transfers (and the
massive quark loops are resolved in the full theory calculation), the
effective theory overestimates the gluon fusion contribution by an
order of magnitude.\footnote{A dedicated comparison of the full matrix
  element with the effective theory is an interesting question in
  itself, which we save for a separate study \cite{future}.}

Due to the particular shape of the re-weighting in
Fig.~\ref{fig:re-weight} we can always find a set of selection cuts
for which effective theory and full calculation agree at the cross
section level. Such an agreement, however, is purely accidental as it
trades off a suppression against an excess in two distinct phase space
regions. An effective field theoretic treatment of $hhjj$ production
without performing the described re-weighting must never be trusted
for neither inclusive nor more exclusive analyses.  
%\smallskip

In the hadron-level analysis we cluster jets from the final state
using {\sc{FastJet}} \cite{Cacciari:2011ma} with $R=0.4$ and $p_T\geq
25$~GeV and $|\eta_j|\leq 4.5$, and require at least two jets. We
double $b$ tag the event (70\% acceptance, 1\% fake) and require the
invariant mass of the $b$ jets to lie within 15 GeV of the Higgs mass
of 125 GeV.

To keep matters transparent in the context of the highly involved
$h\to \tau^+\tau^-$ reconstruction, we assume a perfect efficiency of
1 for demonstration purposes throughout.\footnote{We find the tau
  leptons to be rather hard, which can be used to trigger the event
  via the two tau trigger with little signal loss.}  We ask for two
tau leptons that reproduce the Higgs mass of 125 GeV within $\pm 25$
GeV. The precise efficiencies for leptons in the busy hadronic
environment of the considered process at a 14 TeV high luminosity are
currently unknown, but we expect signal and background to be affect in
similar fashion. We remind the reader that no additional requirements
on missing energy or $m_{\text{T2}}$ are imposed, which are known to
reconcile a smaller $\tau$ efficiency in the overall $S/B$~\cite{us2}.

The $b$ jets are removed from the event and jets that overlap with the
above taus are not considered either. We require at least two
additional jets which are termed ``tagging jets'' of the $hhjj$ event.
\vskip1\baselineskip

%\section{Results}
\noindent \underline{\emph{Results.}}  The cut flow of the outlined
analysis can be found in Tab.~\ref{tab:analysis}. There we also
include analyses of signal samples with changed trilinear and
$VV^\dagger hh$ couplings. The latter modifications have to be
interpreted with caution: The $VV^\dagger hh$ couplings are purely
electroweak and identical to the couplings of two Goldstone bosons to
two gauge bosons. In the high energy limit the Goldstone equivalence
theorem tells us that a modification of $VV^\dagger hh$ away from its
SM value is tantamount to unitarity violation, which explains the
large growth of the WBF component for $\zeta\neq 1$ (such an issue is
not present for $\xi\neq 1$ even though the electroweak sector is
ill-defined). The energy dependence of the matrix element is
effectively cut-off by the parametric Bjorken-$x$ suppression of the
parton distribution functions in the hadronic cross section. In models
in which unitarising degrees are non-perturbative such a behavior is
expected at least qualitatively. We leave an in depth theoretical
discussion on approaches to parameterising such coupling deviations to
future accords.

As can be seen from Tab.~\ref{tab:analysis}, the $hhjj$ analysis in
the $b\bar b \tau^+ \tau^- jj$ channel will be challenging. However,
we remind the reader that no additional selection criteria have been
employed that are known to improve $S/B$ in ``ordinary'' $hh\to b\bar
b \tau^+\tau^-$ analysis~\cite{us,us2}. The arguably straightforward
strategy documented in Tab.~\ref{tab:analysis} should rather be
considered establishing a baseline for a more exhaustive
investigation~\cite{future} than the final verdict on $pp\to hhjj+X$
production.

The gluon fusion contribution dominates the signal component in the
signal region, rendering the WBF contribution almost completely
negligible for analysis with standard $VV^\dagger hh$ coupling
choices. The behaviour of the cross section as function of the Higgs
trilinear interaction results from destructive interference as is
anticipated for studies in $pp\to hh+X$~\cite{us,hhjj2}.

With only about 30 expected WBF events in 3/ab, there is little
leverage in the invariant dijet mass distribution to purify the
selection towards WBF without jeopardising statistical power. On the
other hand, depending on the mechanism of electroweak symmetry
breaking, a large enhancement of the WBF contribution can outrun the
dominant GF events. On a more positive note, if a trilinear Higgs
coupling measurement is obtained from other channels such as $pp\to
hh+X$, this information can in principle be used in the above analysis
to obtain a confidence level interval for the quartic Higgs-gauge
couplings in a simple hypothesis test.

A dedicated analysis which employs techniques motivated recently for
di-Higgs final states~\cite{us2}, as well as methods to separate WBF
from GF based on energy momentum flow observables and kinematic
information~\cite{Englert:2013opa}, jet substructure~\cite{us}, and/or
matrix elements~\cite{Andersen:2012kn}, is likely to significantly
enhance $S/B$, especially when additional limiting factors of $b$ and
tau tagging, smearing and trigger issues are treated more
realistically. We are optimistic that this will eventually allow us to
not only add $pp\to hhjj+X$ to the list of available di-Higgs final
states even for a more realistic treatment of trigger issues, and $b$
and tau tagging efficiencies, but also provide an additional handle to
measure Higgs trilinear and quartic Higgs-gauge couplings at a high
luminosity LHC.  \vskip1\baselineskip

%\section{Summary, Conclusions and Outlook}
\noindent \underline{\emph{Summary, Conclusions and Outlook.}} 
Part of the electroweak physics agenda after the late Higgs
discovery will be to phenomenologically reconstruct the symmetry
breaking potential, as well as to precisely unravel the new particle's
role in TeV scale physics. Measurements of the Higgs trilinear and the
quartic Higgs-gauge couplings are highly sensitive parameters in this
context as they provide a clear picture of the Higgs sector dynamics
and an independent cross-check of mechanism that enforces unitarity.

A lot of theoretical work has been devoted to the WBF contribution to
$pp\to hhjj+X$, which is highly interesting for specifically these
reasons and technically rather straightforward. Nonetheless, hardly
any knowledge about the phenomenological relevance of this
contribution as part of a full hadron-level analysis (including other
signal sources as well as background estimate) has been gathered so
far.

This letter summarises the beginning of a program which seeks to
change this. We have presented a first complete and coherent
phenomenological analysis of di-Higgs production in association with
two jets.  Exploiting the full bandwidth of state-of-the-art Monte
Carlo tools, we have focused on what is probably the
phenomenologically most attractive final state in terms of
reconstruction potential, combinatorial limitations, relatively high
signal yield and comparably large background rejection as a first step
towards a more dedicated analysis. Indeed we find that WBF plays a
completely subdominant role compared to GF, with little statistical
handle to change this with traditional techniques even at high
luminosity.

Also, we have showed that, independent of the particular phase space
region that dedicated analysis targets, a reliable modelling of the
signal crucially depends on the realistic generation of the gluon
fusion signal contribution. Gluon fusion must not be based on
effective field theory methods without applying a proper
fully-differential correction procedure. To this end we have developed
a stand-alone library based on {\sc{MadGraph}} and {\sc{GoSam}}, that
implements a phase space point-dependent re-weighting procedure of the
effective theory calculation, keeping all top and bottom mass
dependencies.

The results indicate that such an analysis at the LHC will be
challenging but not hopeless. In particular, recent developments in
the context of multi-Higgs production have not been exploited in the
present article. We leave this to future work~\cite{future}.
\vskip 0.5 \baselineskip

{\emph{Acknowledgements.}} We thank the organisers of 2013 Les Houches
workshop for convincing us of the relevance and necessity of the
presented work. MJD, CE, and MS thank Margarete M\"uhlleitner and
Michael Spira for interesting conversations during Les Houches
2013. MJD, CE and MS also thank Mike Johnson, Peter Richardson, and
Ewan Steele for computing support. NG wants to thank the other members
of the GoSam collaboration for various useful discussions.
CE is supported by the Institute for Particle Physics Phenomenology
Associateship programme.

%%%%%%%%%%%%%%%%%%%%%%%%%%%%%%%%%%%%%%%%%%%%%%%%%%%%%%%%%%

\end{document}